# Filtering electrons by mode coupling in finite semiconductor superlattices

Xiaoguang Luo[1,2,*], Jian Shi[1], Yaoming Zhang[1,3], Ziang Niu[1,3], Dongpeng Miao[1], Huiru Mi[1], and Wei Huang[1,*]

[1]Frontiers Science Center for Flexible Electronics (FSCFE), Shaanxi Institute of Flexible Electronics (SIFE) & Shaanxi Institute of Biomedical Materials and Engineering (SIBME), Northwestern Polytechnical University, 127 West Youyi Road, Xi'an, 710072, China.
[2]Laboratory of Solid State Microstructures, Nanjing University, Nanjing 210093, China.
[3]School of Materials Science and Engineering, Northwestern Polytechnical University, Xi'an, 710129, China.
*iamxgluo@nwpu.edu.cn (X.L.); iamwhuang@nwpu.edu.cn (W.H.)

Electron transmission through semiconductor superlattices is studied with transfer matrix method and resonance theory. The formation of electron band-pass transmission is ascribed to the coupling of different modes in those semiconductor superlattices with the symmetric unit cell. Upon Fabry-Pérot resonance condition, Bloch modes and two other resonant modes are identified to be related to the nature of the superlattice and its unit cell, respectively. The bands related to the unit cell and the superlattice overlap spontaneously in the tunneling region due to the shared wells, and the coupling of perfectly resonances results in the band-pass tunneling. Our findings provide a promising way to study electronic systems with more complicated superlattices or even optical systems with photonic crystals.

Electronic band structures of semiconductors are reliably manipulated with their nanostructured counterparts because of the quantum confinement effect, such as quantum dots,[1, 2] nanowires,[3] two-dimensional materials,[4] superlattices,[5] etc. Among these structures, semiconductor superlattices have attracted great attention since their carrier transmission behavior is firstly studied by Esaki and Tsu.[6, 7] Ascribe to the periodic nature, the fundamental physical mechanisms of carrier transmission inside superlattices can be readily revealed by the basic theoretical and experimental techniques of solid-state physics.[8-10] In the ballistic regime, the coherent resonant transmission might appear. Nevertheless, the resonances, e.g., Fabry-Pérot (FP) resonances, destroy the band-pass transmission which plays an important role in high-efficiency quantum cascade lasers,[11, 12] high-performance thermoelectric power generators,[13-17] etc. Some non-conventional superlattices have been proposed to realize the band-pass transmission. Tung and Lee[18] designed a band-pass filter in theory by introducing the Gaussian potential profile of segments to suppress the FP oscillation. From the experimental point of view, Gómez et al.[19] modified the model by considering the Gaussian potential profile of barriers only, which was then verified with GaAs/Al$_x$Ga$_{1-x}$As superlattices in experiments.[20] In addition, Gaussian superlattices can also be formed by electrostatic gating.[21] However, the doping level of barriers and the mass of nanoscale electrodes challenge the experiments of non-conventional superlattices. Conventional superlattices can also display the band-pass transmission. With the antireflection coating, Pacher et al.[22] obtained an increase of the transmission through GaAs/AlGaAs superlattice miniband, which was ascribed to the formed FP cavity. This effective design for band-pass transmission was successfully applied to several systems,[15, 16, 23, 24] although the basic mechanism was still obscure. In this paper, we aim to uncover the fundamental physics of band-pass transmission of coated superlattices, which is still periodic if choosing a symmetric unit cell. With the transfer matrix techniques, it is found that the band-pass transmission stems from the coupling of different modes (related to barriers, the symmetric unit cell, and the global superlattice). Moreover, our findings can be used to design band-pass filters by arbitrary periodic superlattices with symmetric unit cell.

## Results and discussion

**Transfer matrix method for electron transmission.** Electrons (similar for holes) in a uniform conductor might be transmitted without any collision when the travel distance is shorter than the mean free path. The electron wavefunction $\psi(x)$ in a one-dimensional (1D) conductor, as sketched in Fig. 1(a), can be described by the stationary Schrödinger equation under the effective mass approximation:[25] $\left[-\frac{\hbar^2}{2}\frac{d}{dx}\frac{1}{m^*(x)}\frac{d}{dx} + V(x)\right]\psi(x) = E\psi(x)$, where $\hbar$ is the reduced Planck constant, $E$ is the electron energy, and $m^*$ and $V$ are the effective mass of electron and potential, respectively. The solution is the forward traveling wave $Ae^{-ikx}$ or the backward traveling one $Be^{ikx}$ due to the absence of reflection, where the wavevector is $k = \sqrt{2m^*(x) \times [E - V(x)]}/\hbar$.

The reflection should be considered in a 1D superlattice or other nonuniform conductors. For the uniform or approximately uniform segment inside nonuniform conductors, the wavefunction can be expressed as $\psi_j(x) = A_j e^{ik_j x} + B_j e^{-ik_j x}$ at the $j$th uniform segment (width $d_j$). Combined with the continuity conditions of wavefunction $\psi(x)$ and the probability current density of electron $\frac{1}{m_j^*}\frac{d\psi(x)}{dx}$, the wave travelling through the $j$th uniform segment and into the $(j + 1)$th uniform segment can be described by matrices of[16]

$$P_j = \begin{pmatrix} e^{-ik_j d_j} & 0 \\ 0 & e^{ik_j d_j} \end{pmatrix} \tag{1a}$$

and

$$T_{j,j+1} = \frac{1}{2}\begin{pmatrix} 1 + \frac{k_{j+1}m_j^*}{k_j m_{j+1}^*} & 1 - \frac{k_{j+1}m_j^*}{k_j m_{j+1}^*} \\ 1 - \frac{k_{j+1}m_j^*}{k_j m_{j+1}^*} & 1 + \frac{k_{j+1}m_j^*}{k_j m_{j+1}^*} \end{pmatrix} \quad (1b)$$

respectively. Thus, the transfer matrix of the whole structure is $M = \cdots T_{j-1,j} \cdot P_j \cdot T_{j,j+1} P_{j+1} \cdots$. The transmission probability is calculated by $\mathcal{T}_{1,N} = \frac{k_N m_1^*}{k_1 m_N^*}\left|\frac{1}{M_{11}}\right|^2$ and the total phase shift (PS) $\varphi$ equals the phase of $1/M_{11}$, where the $k_{1/N}$ and $m_{1/N}^*$ are related parameters of outmost segments. For some simplified potential structures, e.g., Kronig-Penney superlattice, electrons with certain energies can be 100% transmitted, and the PS satisfies the FP resonance condition:[26]

$$\varphi = z\pi \ (z = 1,2,3,\cdots) \quad (2)$$

Resonances at such metastable states are produced by coherent interference.

We next choose the 1D InP/InAs semiconductor heterostructures for further calculations, with the parameters of $m_{InAs}^* = 0.023m_0$, $m_{InP}^* = 0.08m_0$,[27] and the barrier height of 0.57 eV,[28] where $m_0$ is the electron mass in vacuum. InP/InAs heterostructures have been epitaxially grown and embedded in InAs nanowires with the interface abruptness on the level of monolayers.[28-30] For convenience, the parameters of InAs and InP are distinguished with subscripts of "$w$" and "$b$", respectively, in the following. Owing to the tunability of Fermi level, potentials of two different materials are considered as $V_w = 0$ eV and $V_b = 0.57$ eV. Generally, FP resonances can be easily generated in a finite uniform region. For the InP single-barrier embedded in InAs nanowire, the matrix element $M_{11} = \cos(k_b d_b) - i(k_w m_b/k_b m_w + k_b m_w/k_w m_b)\sin(k_b d_b)/2$, and one can get an analytical expression of the transmission probability $\mathcal{T} = 1/[1 + (k_w m_b/k_b m_w - k_b m_w/k_w m_b)^2 \sin^2(k_b d_b)/4]$. Some perfectly coherent resonances at $\mathcal{T} = 1$ can be found with the energy levels $E_{n=1,2,3,\cdots}^*$. Based on the FP resonance condition, PSs at the metastable states satisfy $k_b(E_n^*) \cdot d_b = n\pi$, indicating $k_b(E_n^*)$ is a real wavevector and $E_n^* > V_b$, as sketched in Fig. 1(b). When $d_b = 2.5$ nm, the transmission spectrum and the corresponding PS have been calculated by transfer matrix technique and shown in Fig. 1(c). Two peaks appear in the region of $E < 5$ eV, and PSs are $\pi$ and $2\pi$, respectively, confirming the energy values of two peaks are definitely $E_1^*$ and $E_2^*$.

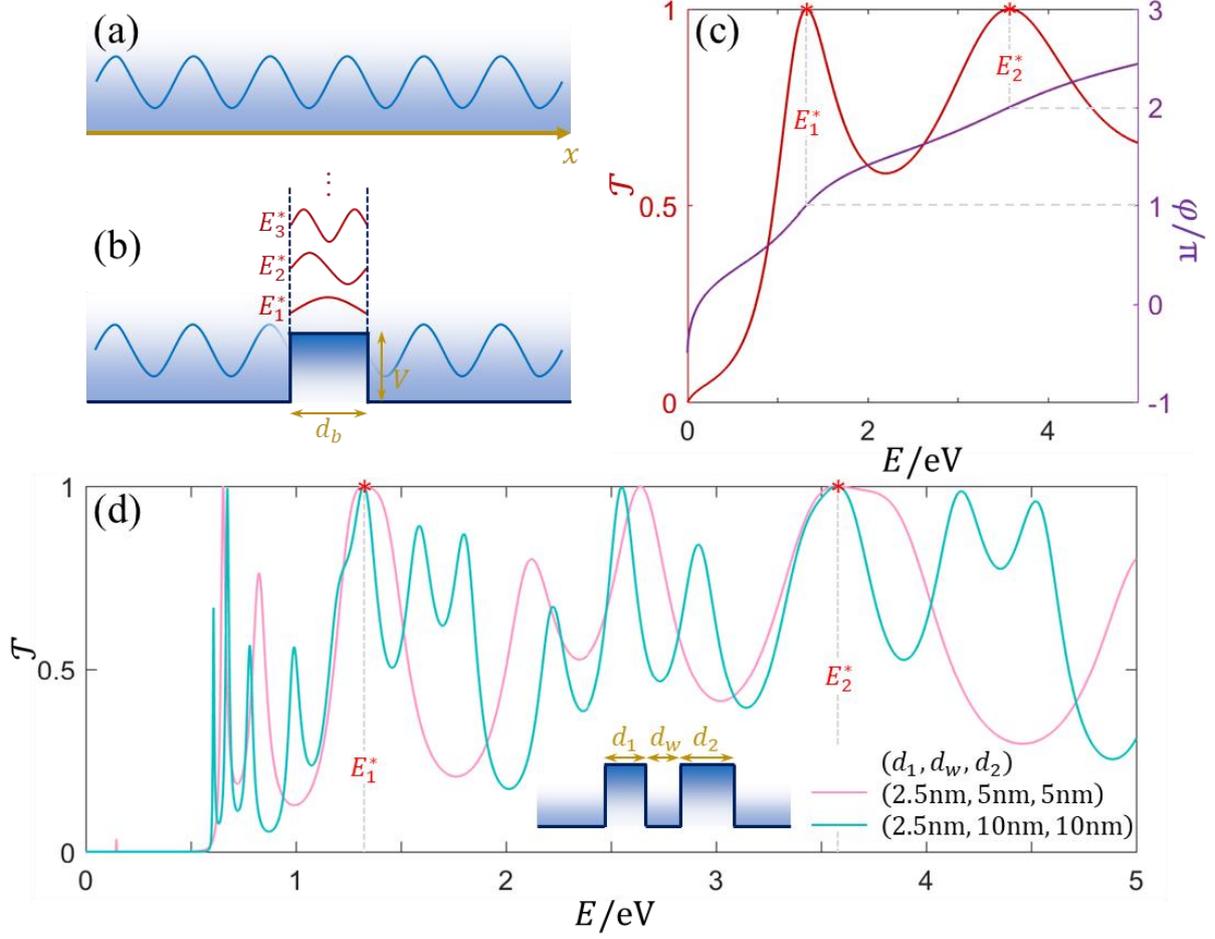

Fig. 1 (a) Electronic wave travels in a uniform conductor. (b) Electron transmission through a single-barrier with the barrier height $V$ and barrier width $d_b$, the electronic wave at energy of $E_n^*$ denotes the $n$th resonant mode, whose PS is $n\pi$. (c) The transmission spectrum and PS of the InP single-barrier when $d_b = 2.5$ nm, peaks at $E_1^*$ and $E_2^*$ correspond to the resonant modes. (d) The transmission probability of the InP/InAs/InP double-barrier with different widths. Inset is the potential profile of the asymmetric double-barrier structure.

**Resonant Modes classification.** Besides the single-barrier, similar resonant transmission can also happen in multiple-barrier structures. If $d_\mathrm{I}$ is one common divisor length of all barrier widths $(d_1, d_2, d_3, \cdots)$ of the structure, the metastable states at energy $E_n^*$ in the single-barrier with width $d_\mathrm{I}$ will also appear in the multiple-barrier structure, because of the severalfold $\pi$ PS inside all barriers. The transmission spectrum of InP/InAs/InP double-barrier has been calculated after fixing the first barrier width (i.e., $d_1 = 2.5$ nm). The second barrier width $d_2$ and the well width $d_w$ impact the transmission significantly, and more different resonances can be seen in Fig. 1(d). However, resonances arising in the single-barrier with width $d_1 = 2.5$ nm is robust even at the double-barrier structure when the second barrier width is several times of $d_1$, no matter what the well width is chosen, as shown the unity peaks at $E_1^*$ and $E_2^*$. For convenience, this kind of modes originated from barriers are called *bulk modes* in the following.

The enlargement of Fig. 1(d) shows no other unity peak, and the minimum PS of all peaks is more than $3\pi$ for $d_2 = 5$ nm and more than $5\pi$ for $d_2 = 10$ nm. Similar results can be recovered in other asymmetric multiple-barrier structures. The FP resonance condition is invalid even for the perfectly coherent resonances at the asymmetric structure. To uncover the reason, the simplified symmetric situation is investigated, i.e., $d_1 = d_2$. The widths of each InP barrier and InAs well are chosen as $d_b = 2.5$ nm and $d_w = 5$ nm in the following, unless otherwise stated. Fig. 2(a) displays the transmission spectrum of the symmetric double-barrier structure when $E < 5$ eV. Different from the asymmetric structure, all peaks are unity in the transmission spectrum. Each passband contains one or two peaks, implying modes coupling in the transmission. Different peaks are easily used to distinguish those modes. After compared with Fig. 1, peaks in the two-peak passbands are identified as the resonant modes of 2.5 nm single-barrier, as labeled $E_1^*$ and $E_2^*$ in Fig. 2(a). Other peaks at $E_{n=1,2,3,4,5}$ should be determined by the whole potential structure. However, calculated results in Fig. 2(b) show the PS $\varphi$ deviates the resonance condition at $E_{n=1,2,3,4,5}$. Inspired from the uniform conductor, whose crystal lattice is periodic at the atomic scale, the symmetric double-barrier here can also be reconfigured to a periodic structure after adding an extra well at one side, as shown in the dashed box in the Inset of Fig. 2(a). So far, the symmetric InP/InAs/InP double-barrier structure can be regarded as 2-period superlattice with the unit cell of InP/InAs. Then the PS at $E_{n=1,2,3,4,5}$ is $(2n - 1)\pi$, as shown in Fig. 2(b).

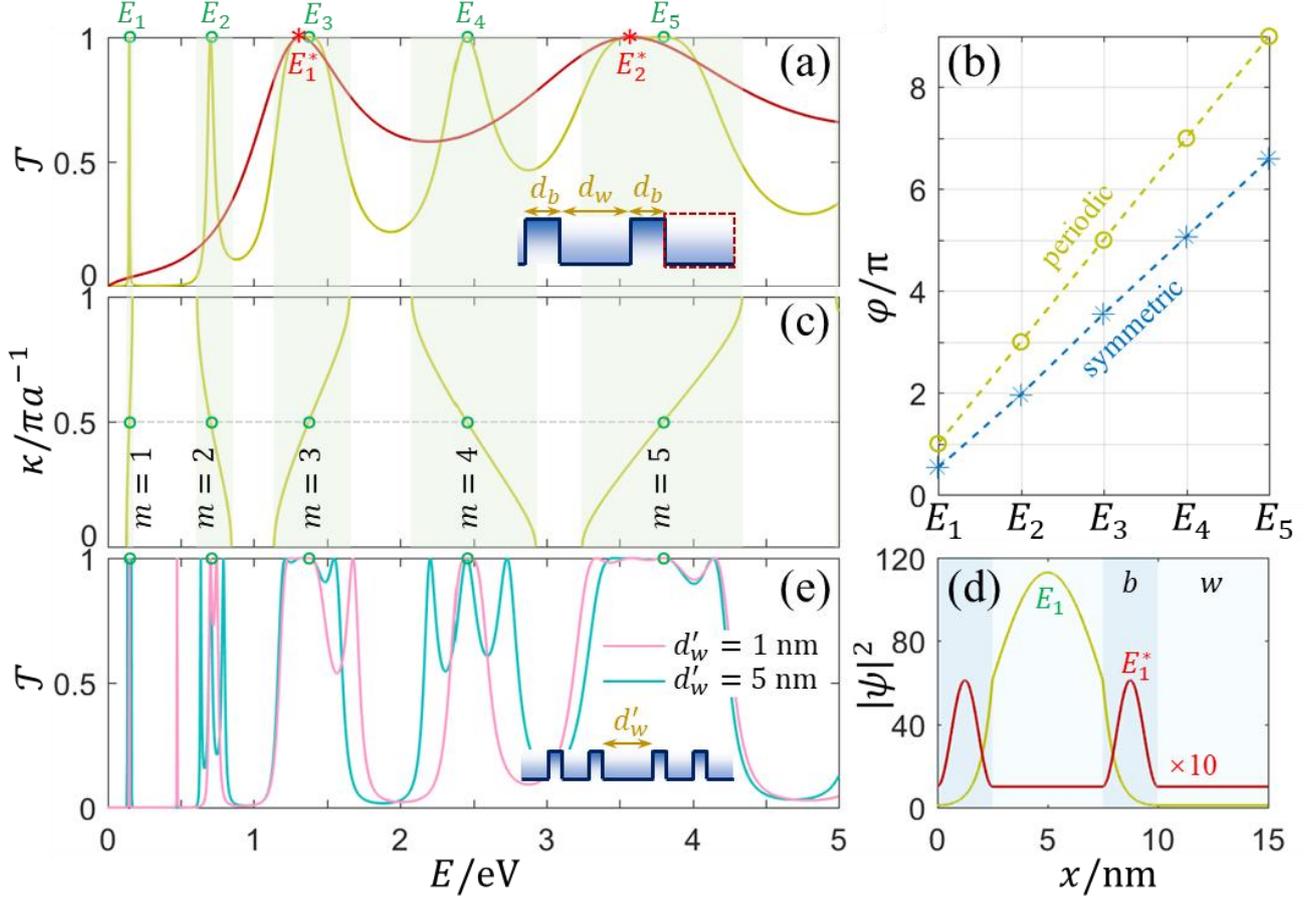

Fig. 2 (a) The transmission spectrum of the symmetric double-barrier structure when $d_b = 2.5$ nm and $d_w = 5$ nm, the red line is the result of 2.5 nm single-barrier. Inset is the potential profile of the symmetric double-barrier structure. (b) The PSs for the symmetric structure and the reconfigured periodic structure. (c) Band structure of the infinite superlattice, where $m$ denotes the $m$th miniband (or Brillouin zone). (d) $|\psi|^2$ spatial distribution of two different resonances at $E_1$ and $E_1^*$, respectively. (e) The transmission spectra of two double-barrier structures connected with InAs well at the width of $d_w'$. Insets in (a) and (e) are the related potential profiles, and circles and stars in (a), (c), and (e) mark peaks in the transmission spectra.

The electronic bands of the infinite superlattice can be calculated with the Bloch theorem

$$M_u \psi = e^{i\kappa a}\psi \qquad (3)$$

where $M_u$ is the transfer matrix of the unit cell, $a$ is the lattice constant, and $\kappa$ is the Bloch wavevector. The simplification of Eq. (3) for InP/InAs superlattice is $\cos(\kappa a) = \cos(k_b b)\cos(k_w w) - (k_w m_b/k_b m_w + k_b m_w/k_w m_b)\sin(k_b b)\sin(k_w w)/2$. Fig. 2(c) shows the band structure, where allowed bands are consistent with the transmission spectrum. Notice that not all states at each miniband are perfect resonances for finite superlattice. The FP resonance condition enable the determination of those discrete metastable states of perfect resonances. It is known that the Bloch wavevector $\kappa$ can be regarded as an effective vector in a periodic conductor, so that the PS of the discrete state in the $N$-period superlattice is expressed as $\varphi = N\kappa a$. At the $m$th Brillouin zone [i.e., $(m-1)\pi/a < \kappa < m\pi/a$], the value of $\kappa = (m - 1 + n/N)\pi/a$ holds the resonance condition,[31-35] where $n =$ $1,2,3,\cdots,N-1$ is the resonance index in $m$th Brillouin zone, namely there are $N-1$ resonances in each Brillouin zone. For our 2-period InP/InAs superlattice, the intersection points of $\kappa = \pi/2a$ and the dispersion curves in Fig. 2(c) correspond with the resonance energy levels of $E_{n=1,2,3,4,5}$. These modes originated from the periodicity can be called as *Bloch modes*.[36]

The added well for reconfigured periodic structure has no impact on the transmission due to the same value of $|M_{11}|^2$ and $|M_{11}e^{\pm ik_w d_w}|^2$. Combined with the finite element method, the wavefunction $\psi$ can be solved with respect to $x$. For the 2-period InP/InAs superlattice, $|\psi|^2$ of two different resonant modes at $E_1$ and $E_1^*$ are shown in Fig. 2(d). $|\psi|^2$ is distributed in barriers for $E_1^*$ and in the symmetric double-barrier structure for $E_1$, confirming the PS complement effect of the added well. Transmission modes related to $E_{n=1,2,3,4,5}$ in the symmetric structure are similar with those bulk modes in single-barrier, except the counting way of PS. This deduction can be confirmed by two symmetric double-barrier structures connected with an extra InAs well (width $d_w'$). The transmission spectra of this structure are shown in Fig. 2(e) when $d_w' = 1$ nm and $d_w' = 5$ nm, where the robust resonant modes of one double-barrier still appear in the transmission spectra. For convenience, this kind of modes are called *symmetric modes* in the following and are equivalent to the Bloch modes if the structure becomes periodic after considering the added well.

The band structure of the InP/InAs superlattice is obtained by $\text{Det}(P_b T_{b,w} P_w T_{w,b} - e^{i\kappa a}) = 0$, which can be transferred to $\text{Det}(P_{b2} T_{b,w} P_w T_{w,b} P_{b1} - e^{i\kappa a}) = 0$ if $d_{b1} + d_{b2} + d_w = a$. In other words, the unit cell of the periodic superlattice can be arbitrarily chosen with the invariable band structure, as sketched by the dashed boxes in Fig. 3(a). Fig. 3(c) shows the band structure of the InP/InAs infinite superlattice with the InP barrier width of 5 nm. The choice of unit cell significantly impacts the transmission of the superlattice with the finite periodicity $N$. Figs. 3(d) and 3(f) show the transmission spectrum of superlattice with the unit cell of InP/InAs structure when $N = 10$. It is found that nine or ten unity peaks in each Brillouin zone. Based on the analysis of Fig. 2, nine unity peaks come from Bloch (or symmetric here) resonant modes, and the extra one peak (in accord with resonances of a 5 nm barrier) in some Brillouin zones is the bulk resonant mode. For the unit cell of the asymmetric InP/InAs/InP structure (i.e., $d_{b1} \neq d_{b2}$), nine Bloch resonant modes reappear in each Brillouin zone (not shown here). For the unit cell of symmetric InP/InAs/InP structure (i.e., $d_{b1} = d_{b2}$), some more interesting unity peaks arise. The main cause here is that the symmetric unit cell itself can also be regarded as the 2-period local superlattice with a local InP/InAs unit cell (for comparison, the symmetric InP/InAs/InP structure is the unit cell of the global superlattice). The band structure of the local superlattice with infinite periodicity is shown in Fig. 3(c). The transmission spectrum when $N = 10$ is shown in Figs. 3(e) and 3(g), implying that the band structure of the global superlattice determines allowed or forbidden bands. Therefore, energy levels of $E_{n=1,2,3,4,5}$ at $\kappa = \pi/2a$ from dispersion curves of the local superlattice are the resonance energies only if they are located inside the allowed bands. Besides Bloch resonances of global superlattice, one extra Bloch resonance (as remarked by green circles in Fig. 3) is in 1st, 2nd, 4th, 6th, and 7th Brillouin zone when $E < 5$ eV. In addition, one more bulk resonance is found in 4th and 7th Brillouin zone (as remarked by red stars in Figs. 2 and 3). Consequently, the transmission spectrum of the $N$-period superlattice with the unit cell of symmetric InP/InAs/InP structure is related to the coupling between Bloch modes, symmetric modes, and bulk modes.

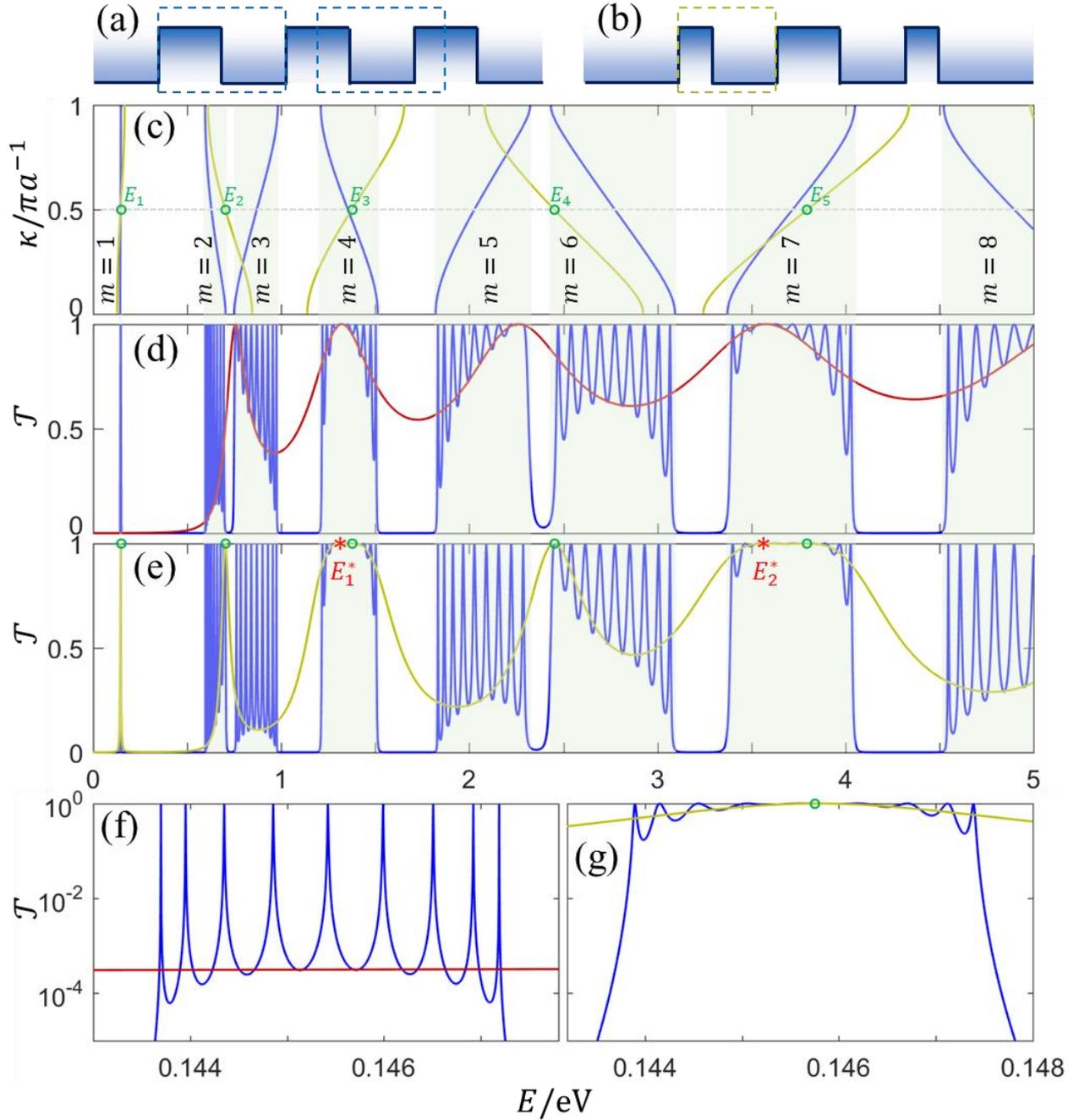

Fig. 3 (a) The potential profile of periodic InP/InAs superlattice, dashed boxes mark the unit cells. (b) The potential profile of 2-period symmetric InP/InAs/InP structure. (c) Band structures of periodic InP/InAs infinite superlattices when the InP barrier width is $2d_b = 5$ nm (blue line) and $d_b = 2.5$ nm (doderblue line), respectively. (d) Transmission spectra for the 10-period InP/InAs superlattice (blue line) and for the single-barrier structure (red line), and barrier width is $2d_b$. (e) Transmission spectra for 10-period superlattice with the unit cell of the symmetric InP/InAs/InP structure (blue line) and for the symmetric InP/InAs/InP structure only (doderblue line), and barrier width is $d_b$. (f) and (g) are enlargements of (d) and (e), respectively. All well widths are $d_w = 5$ nm. Circles and stars mark peaks in transmission spectra.

**Transmission by mode coupling.** The coupling of multiple modes significantly improves the electron transmission in allowed bands. Theoretically, there are $N-1$ unity transmission peaks in each Brillouin zone ascribed to Bloch modes because of the wave interference, as well as $N-2$ deeps with transmission probability $T=0$ between neighboring peaks. However, zero deeps are never reached due to the modes coupling. As shown in Figs. 3(f) and 3(g), deeps are improved to the orders of ~$10^{-4}$ and ~$10^0$ in the first Brillouin zone (i.e., the tunneling region here) for two different superlattices, which are close to the transmission probabilities of the unit cell itself, especially around the symmetric resonances. The empirical explanation of the improvement might be the independent transmission of three different modes during coupling. At the first Brillouin zone, the coupling of Bloch modes and symmetric modes dominates the transmission due to the perfectly coherent resonance. By carefully choosing barrier and well widths, the perfectly coherent resonance energies of the local modes ($E^*_{n=1,2,3,\cdots}$ and $E_{n=1,2,3,\cdots}$) can be tuned to the middle of the allowed band, and the probabilities in arbitrary Brillouin zone can then be improved to a high level, finally forming the band-pass transmission.

It is found that symmetric resonances always arise in the allowed bands of the tunneling region (where $E<V$), which can be ascribed to the shared wells of global and local superlattices. For $E>V$, barriers of unit cells become important roles in PS accumulation because of the real wavevector, resulting in different overlap of bands, as shown in Fig. 3(c). To confirm and extend the deduction, 3-period superlattice with the unit cell of symmetric InP/InAs/InP/InAs/InP triple-barrier is studied by the same method. The band structures of global and local superlattices with infinite periodicity are shown in Fig. 4(a), where two allowed minibands appear in the tunneling region. Based on the bands of local superlattice, the perfectly coherent resonances take place at $\kappa=\pi/3a$ and $\kappa=2\pi/3a$ for finite periodicity, which are indeed located inside the 1st and 2nd allowed minibands, as labeled $E_1$ and $E_2$ in Fig. 4(a). Deeps in the transmission spectrum are improved to the level of $10^0$, as shown in Fig. 4(b). This band-pass transmission is from the coupling of Bloch modes, symmetric modes, and bulk modes (although with a very small contribution). In fact, the symmetric unit cell has no need to be periodic, i.e., barrier widths can be different, and accordingly the added well (for PS complement) might be unique for each local resonance. Fig. 4(c) shows the transmission spectrum when the middle barrier width $d'_b=3$ nm of the symmetric InP/InAs/InP/InAs/InP unit cell. Similar coupling and improvement can also be obtained in arbitrary superlattice (not limited at Kronig-Penney superlattice) with symmetric unit cell, while the width choice of each segment should support perfect resonances. Besides the one-dimensional nanowire superlattice, the two- or three-dimensional semiconductor superlattice can also be studied by employing our approach, such as the in-plane superlattice or out-plane superlattice of two-dimensional materials[37, 38]. Therein, the incidence angle and parameter components along the superlattice are critical to the further interpretation.

Noticed that resonances and modes coupling of our superlattice can also be interpreted by the Kard parametrization that in terms of both PS and impedance[39]. To verify the validity of our interpretation, the theory of quantum-mechanical wave impedance (QMWI)[18, 40] is introduced to calculated the 3-period superlattice again. The characteristic impedance of $j$th uniform segment (width $d_j$) is defined as

$$Z_{j,0} = \frac{2\hbar k_j}{m_j} \tag{4}$$

and the impedance at the left side of $j$th segment can be calculated by using

$$Z_j = Z_{j,0} \frac{Z_{j+1}\cosh(ik_j d_j) - Z_{j,0}\sinh(ik_j d_j)}{Z_{j,0}\cosh(ik_j d_j) - Z_{j+1}\sinh(ik_j d_j)} \tag{5}$$

It should be noted that the impedance of the rightmost segment equals to its characteristic impedance because of no potential discontinuity any more. After repeatedly calculating with Eq. (5), the impedance of the whole superlattice can be obtained, written as $Z_S$, which can be regarded as the load impedance of the whole superlattice. Thus, the transmission probability is given by

$$\mathcal{T} = 1 - |t|^2 \tag{6}$$

where the reflection coefficient $t = (Z_S - Z_{1,0})/(Z_S + Z_{1,0})$. The calculated results of the 3-period superlattice are shown in Figs. 4(b-e). It is found that the transmission probabilities obtained from two different approaches agree well with each other. In the band-pass region, the real part of the load impedance $Z_S$ is close to the characteristic impedance $Z_{1,0}$, and the imagine part tends to zero, indicating the good impedance matching for the band-pass transmission.

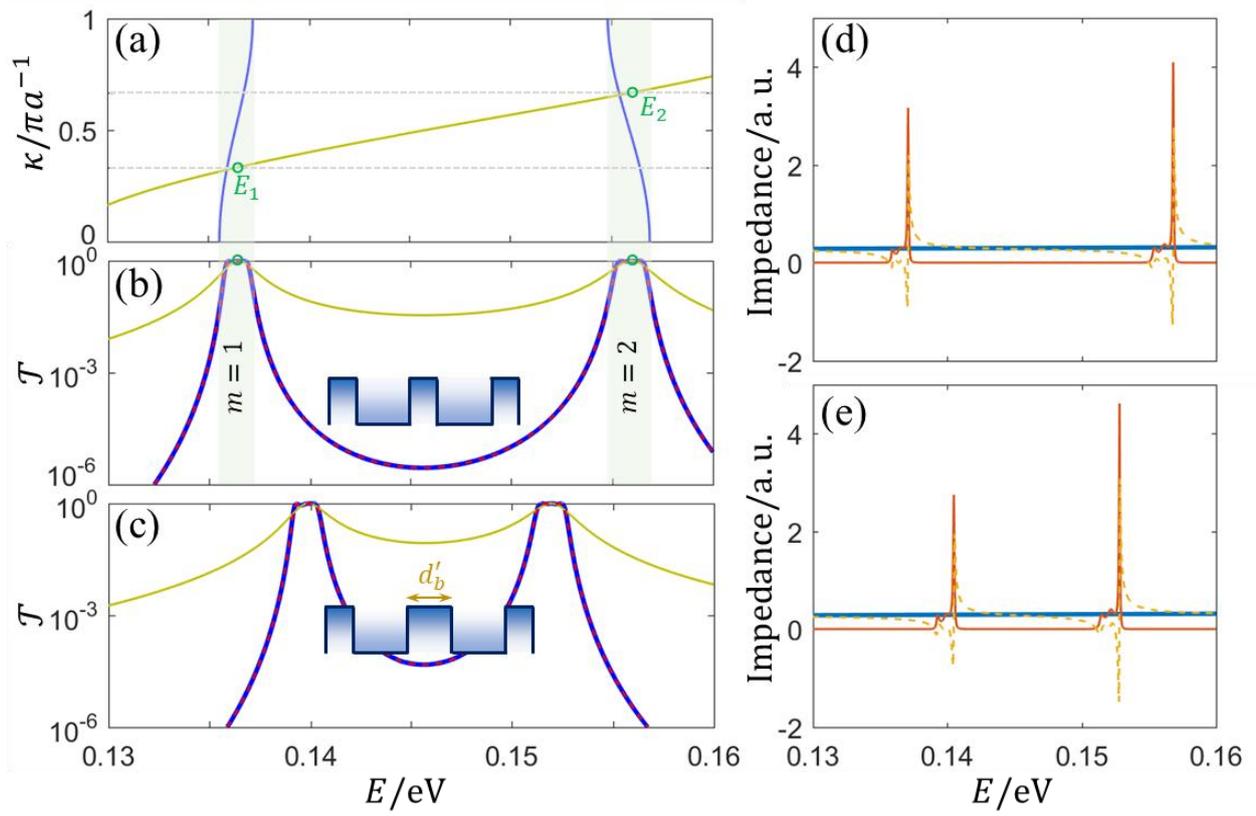

Fig. 4 (a) Band structures of periodic triple-barrier (InP/InAs/InP/InAs/InP) infinite superlattice (blue line) and periodic single-barrier (InP/InAs) infinite superlattice (doderblue line), where $d_b = 2.5$ nm and $d_w = 5$ nm. (b) Transmission spectra for 3-period superlattice with the unit cell of periodic triple-barrier (blue line) and for the unit cell itself (doderblue line), the dashed red line indicates the results from the theory of QMWI. Circles mark peaks in transmission spectra of unit cells. (c) Transmission spectra of 3-period superlattice with symmetric triple-barrier unit cell when the middle barrier width is $d'_b = 3$ nm (blue line), and of the unit cell itself (doderblue line), the dashed red line indicates the results from the theory of QMWI. (d) and (e) are the corresponding impedances of the (b) and (c), respectively, where the bold solid line indicates $Z_{1,0}$, and the fine lines indicate the real parts (solid lines) and imagine parts (dashed lines) of $Z_S$.

## Conclusions

In conclusion, bulk modes, symmetric modes, and Bloch modes are identified through superlattices with the unit cell from single-barrier, double-barrier, to multiple-barrier. The transmission of an arbitrary superlattice could be the coupling result of these modes. The coupling of transmission modes from the unit cell and Bloch modes from the periodicity can remarkably improve the transmission probabilities. Once the unit cell is symmetric, the symmetric modes appear, and the coupling with bulk and Bloch modes might result in the band-pass transmission. In particular, the spontaneous coupling (stem from the shared wells) between symmetric modes and Bloch modes makes the band-pass tunneling, which might be useful for quantum cascade lasers[11, 12] and thermoelectric energy conversion[13-17]. Our interpretation can also be adopted to study electronic systems with more complicated potential profiles and the optical systems with photonic crystals.

# References

1. Ekimov, A. I.; Efros, A. L.; Onushchenko, A. A., Quantum size effect in semiconductor microcrystals. *Solid State Communications* **1985,** *56* (11), 921-924.
2. Reed, M. A.; Bate, R. T.; Bradshaw, K.; Duncan, W. M.; Frensley, W. R.; Lee, J. W.; Shih, H. D., Spatial quantization in GaAs–AlGaAs multiple quantum dots. *Journal of Vacuum Science & Technology B: Microelectronics Processing and Phenomena* **1986,** *4* (1), 358-360.
3. Rurali, R., Colloquium: Structural, electronic, and transport properties of silicon nanowires. *Reviews of Modern Physics* **2010,** *82* (1), 427-449.
4. Novoselov, K. S.; Geim, A. K.; Morozov, S. V.; Jiang, D.; Zhang, Y.; Dubonos, S. V.; Grigorieva, I. V.; Firsov, A. A., Electric Field Effect in Atomically Thin Carbon Films. *Science* **2004,** *306* (5696), 666-669.
5. Esaki, L.; Tsu, R., Superlattice and Negative Differential Conductivity in Semiconductors. *IBM Journal of Research and Development* **1970,** *14* (1), 61-65.
6. Tsu, R.; Esaki, L., Tunneling in a finite superlattice. *Applied Physics Letters* **1973,** *22* (11), 562-564.
7. Chang, L. L.; Esaki, L.; Tsu, R., Resonant tunneling in semiconductor double barriers. *Applied Physics Letters* **1974,** *24* (12), 593-595.
8. Smith, D. L.; Mailhiot, C., Theory of semiconductor superlattice electronic structure. *Reviews of Modern Physics* **1990,** *62* (1), 173-234.
9. Pereyra, P., The Transfer Matrix Method and the Theory of Finite Periodic Systems. From Heterostructures to Superlattices. *Physica Status Solidi (b)* **2022,** *259* (3), 2100405.
10. Ricco, B.; Azbel, M. Y., Physics of resonant tunneling. The one-dimensional double-barrier case. *Physical Review B* **1984,** *29* (4), 1970-1981.
11. Gmachl, C.; Capasso, F.; Sivco, D. L.; Cho, A. Y., Recent progress in quantum cascade lasers and applications. *Reports on Progress in Physics* **2001,** *64* (11), 1533-1601.
12. Scamarcio, G.; Capasso, F.; Sirtori, C.; Faist, J.; Hutchinson Albert, L.; Sivco Deborah, L.; Cho Alfred, Y., High-Power Infrared (8-Micrometer Wavelength) Superlattice Lasers. *Science* **1997,** *276* (5313), 773-776.
13. Whitney, R. S., Most Efficient Quantum Thermoelectric at Finite Power Output. *Physical Review Letters* **2014,** *112* (13), 130601.
14. Luo, X.; Liu, N.; Qiu, T., Efficiency at maximum power of thermochemical engines with near-independent particles. *Physical Review E* **2016,** *93* (3), 032125.
15. Mukherjee, S.; Muralidharan, B., Electronic Fabry-Perot Cavity Engineered Nanoscale Thermoelectric Generators. *Physical Review Applied* **2019,** *12* (2), 024038.
16. Luo, X.; Zhang, H.; Liu, D.; Han, N.; Mei, D.; Xu, J.; Cheng, Y.; Huang, W., Efficiency at maximum power of thermoelectric heat engines with the symmetric semiconductor superlattice. *Physica E: Low-dimensional Systems and Nanostructures* **2021,** *129*, 114657.
17. Karbaschi, H.; Lovén, J.; Courteaut, K.; Wacker, A.; Leijnse, M., Nonlinear thermoelectric efficiency of superlattice-structured nanowires. *Physical Review B* **2016,** *94* (11), 115414.
18. Tung, H.-H.; Lee, C.-P., An energy band-pass filter using superlattice structures. *IEEE Journal of Quantum Electronics* **1996,** *32* (3), 507-512.
19. Gómez, I.; Domínguez-Adame, F.; Diez, E.; Bellani, V., Electron transport across a Gaussian superlattice. *Journal of Applied Physics* **1999,** *85* (7), 3916-3918.
20. Diez, E.; Gómez, I.; Domínguez-Adame, F.; Hey, R.; Bellani, V.; Parravicini, G. B., Gaussian semiconductor superlattices. *Physica E: Low-dimensional Systems and Nanostructures* **2000,** *7* (3), 832-835.
21. Sánchez-Arellano, A.; Madrigal-Melchor, J.; Rodríguez-Vargas, I., Non-conventional graphene superlattices as electron band-pass filters. *Scientific Reports* **2019,** *9* (1), 8759.

22. Pacher, C.; Rauch, C.; Strasser, G.; Gornik, E.; Elsholz, F.; Wacker, A.; Kießlich, G.; Schöll, E., Antireflection coating for miniband transport and Fabry–Pérot resonances in GaAs/AlGaAs superlattices. *Applied Physics Letters* **2001,** *79* (10), 1486-1488.

23. Sharma, A.; Tulapurkar, A. A.; Muralidharan, B., Band-pass Fabry-Pèrot magnetic tunnel junctions. *Applied Physics Letters* **2018,** *112* (19), 192404.

24. Luo, X.; Zhou, M.; Liu, J.; Qiu, T.; Yu, Z., Magneto-optical metamaterials with extraordinarily strong magneto-optical effect. *Applied Physics Letters* **2016,** *108* (13), 131104.

25. Liu, X.-W.; Stamp, A. P., Resonant tunneling and resonance splitting: The inherent properties of superlattices. *Physical Review B* **1994,** *50* (3), 1588-1594.

26. Schiff, L. I., *Quantum Mechanics* McGraw-Hill: New York, 1968.

27. O'Dwyer, M. F.; Humphrey, T. E.; Linke, H., Concept study for a high-efficiency nanowire based thermoelectric. *Nanotechnology* **2006,** *17* (11), S338-S343.

28. Björk, M. T.; Ohlsson, B. J.; Sass, T.; Persson, A. I.; Thelander, C.; Magnusson, M. H.; Deppert, K.; Wallenberg, L. R.; Samuelson, L., One-dimensional Steeplechase for Electrons Realized. *Nano Letters* **2002,** *2* (2), 87-89.

29. Josefsson, M.; Svilans, A.; Burke, A. M.; Hoffmann, E. A.; Fahlvik, S.; Thelander, C.; Leijnse, M.; Linke, H., A quantum-dot heat engine operating close to the thermodynamic efficiency limits. *Nature Nanotechnology* **2018,** *13* (10), 920-924.

30. Prete, D.; Erdman, P. A.; Demontis, V.; Zannier, V.; Ercolani, D.; Sorba, L.; Beltram, F.; Rossella, F.; Taddei, F.; Roddaro, S., Thermoelectric Conversion at 30 K in InAs/InP Nanowire Quantum Dots. *Nano Letters* **2019,** *19* (5), 3033-3039.

31. Pacheco, M.; Claro, F., Simple results for one-dimensional periodic potentials. *Physica Status Solidi (b)* **1982,** *114* (2), 399-403.

32. Vezzetti, D. J.; Cahay, M. M., Transmission resonances in finite, repeated structures. *Journal of Physics D: Applied Physics* **1986,** *19* (4), L53-L55.

33. Kalotas, T. M.; Lee, A. R., One-dimensional quantum interference. *European Journal of Physics* **1991,** *12* (6), 275-282.

34. Sprung, D. W. L.; Wu, H.; Martorell, J., Scattering by a finite periodic potential. *American Journal of Physics* **1993,** *61* (12), 1118-1124.

35. Pereyra, P.; Castillo, E., Theory of finite periodic systems: General expressions and various simple and illustrative examples. *Physical Review B* **2002,** *65* (20), 205120.

36. Sibilia, C.; Benson, T. M.; Marciniak, M.; Szoplik, T., *Photonic Crystals: Physics and Technology*. 1st ed.; Springer: Milano, 2008.

37. Alvarado-Goytia, J. J.; Rodríguez-González, R.; Martínez-Orozco, J. C.; Rodríguez-Vargas, I., Biperiodic superlattices and transparent states in graphene. *Scientific Reports* **2022,** *12* (1), 832.

38. Wang, C.; He, Q.; Halim, U.; Liu, Y.; Zhu, E.; Lin, Z.; Xiao, H.; Duan, X.; Feng, Z.; Cheng, R.; Weiss, N. O.; Ye, G.; Huang, Y.-C.; Wu, H.; Cheng, H.-C.; Shakir, I.; Liao, L.; Chen, X.; Goddard Iii, W. A.; Huang, Y.; Duan, X., Monolayer atomic crystal molecular superlattices. *Nature* **2018,** *555* (7695), 231-236.

39. Sprung, D. W. L.; Vanderspek, L. W. A.; van Dijk, W.; Martorell, J.; Pacher, C., Biperiodic superlattices and the transparent state. *Physical Review B* **2008,** *77* (3), 035333.

40. Khondker, A. N.; Khan, M. R.; Anwar, A. F. M., Transmission line analogy of resonance tunneling phenomena: The generalized impedance concept. *Journal of Applied Physics* **1988,** *63* (10), 5191-5193.

## Acknowledgments

This work was supported by the National Natural Science Foundation of China (Grant No. 61905198), the National Postdoctoral Program for Innovative Talents (No. BX20190283), the Key Research and Development Program of Shaanxi Province (Nos. 2020GXLH-Z-020 and 2020GXLH-Z-027), and the Fundamental Research Funds for the Central Universities.

## Author contributions

X.L. proposed the ideas, carried out the calculation and wrote the main manuscript text. J.S. processed the data and visualized it. X.L., J.S., Y.Z., Z.N., D.M., and H.M. analyzed the results. W.H. provided us with the necessary conditions for the work of this paper. All authors reviewed the manuscript.

## Competing interests

The authors declare no competing interests.

## Additional information

**Correspondence** and requests for materials should be addressed to X.L.